\documentclass[twocolumn,showpacs,preprintnumbers,amsmath,amssymb]{revtex4}

\usepackage{graphicx}
\usepackage{comment}

\begin{document}

\title{Thermalization of a dimerized antiferromagnetic spin chain}

\author{N. P. Konstantinidis}

\affiliation{Max-Planck-Institut f\"ur Physik komplexer Systeme, 01187 Dresden, Germany}

\date{\today}

\begin{abstract}
Thermalization is investigated for the one-dimensional anisotropic antiferromagnetic Heisenberg model with dimerized nearest-neighbor interactions that break integrability. For this purpose the time evolution of local operator expectation values after an interacting quench is calculated directly with the Chebyshev polynomial expansion, and the deviation of the diagonal from the canonical thermal ensemble value is calculated for increasing system size for these operators. The spatial and spin symmetries of the Hamiltonian are taken into account to divide it into symmetry subsectors. The rate of thermalization is found to weaken with the dimerization parameter as the Hamiltonian evolves between two integrable limits, the non-dimerized and the fully dimerized where the chain breaks up into isolated dimers. This conclusion is supported by the distribution of the local operator off-diagonal elements between the eigenstates of the Hamiltonian with respect to their energy difference, which determines the strength of temporal fluctuations. The off-diagonal elements have a low-energy peak for small dimerization which facilitates thermalization, and originates in the reduction of spatial symmetry with respect to the non-dimerized limit. For increasing dimerization their distribution changes and develops a single low-energy maximum that relates to the fully dimerized limit and slows down thermalization.

\end{abstract}

\maketitle

\section{Introduction}
\label{sec:Introduction}
In recent years the field of out-of-equilibrium phenomena in thermally isolated quantum many-body systems has attracted significant attention \cite{Srednicki94,Deutsch91,Cazalilla12,Rigol07,Calabrese11,Reimann08,Rigol08,Fagotti13,Pozsgay13,Sirker14,Rigol09,Rigol09-1,Lazarides14,Lazarides14-1}. A part of an isolated system feels the rest of the system acting as a bath, resulting in probability flowing in and out of the subsystem. This can lead to weakly fluctuating operator expectation values in the subsystem in the long-time limit, which can be described by a statistical ensemble with a small number of parameters. The nature of the statistical ensemble depends on the integrability properties of the Hamiltonian. Unlike the classical case, integrability is not precisely defined in quantum mechanics \cite{Caux11}. The prevailing idea is that in a quantum-integrable system the number of local integrals of motion is proportional to the constituents of the system, something not required at the classical level. A non-integrable Hamiltonian is expected to thermalize according to the canonical thermal ensemble. On the other hand, the conserved quantities of an integrable Hamiltonian have to be taken into account in the generalized Gibbs ensemble that describes correlations in the long-time limit \cite{Rigol08,Rigol07,Fagotti13,Pozsgay13}. It also has been recently shown that the quench action formalism correctly captures the long-time limit steady state \cite{Caux13,Wouters14,Pozsgay14,Brockmann14,Mestyan14}. Hamiltonians where the proximity to integrability is controlled by a parameter are of particular interest, since it is possible to interpolate between the integrable and non-integrable cases in a well-controlled fashion \cite{NPK15,Beugeling15}.

In the case of weak integrability breaking it has been shown that thermalization can be particularly slow. For the fermionic \cite{Kollar11} and the spinless fermionic dimerized Hubbard model \cite{Essler14}, as well as the quantum Ising chain \cite{Marcuzzi13}, local operator expectation values first relax to a non-thermal quasi-stationary value in a prethermalized regime, before eventually relaxing to the canonical thermal ensemble value. Different scenarions of prerelaxation close to integrable points have been considered in \cite{Bertini15}. In the case of the anisotropic antiferromagnetic Heisenberg model (AAHM) with nearest and next-nearest neighbor interactions that break integrability, thermalization is also slow close to the integrable point \cite{NPK15}. This was attributed to the appearance of a low-energy peak for local operator off-diagonal elements that facilitates thermalization once integrability is broken, but which is only weak for small integrability breaking. For stronger integrability breaking the low-energy peak increases in magnitude and thermalization becomes faster. Altogether there are three different thermalization regimes when the next-nearest neighbor interaction is relatively moderate.

In this paper we investigate thermalization for the AAHM for a spin chain with dimerized interactions, where the proximity to integrability is controlled by the dimerization strength. This is defined as the deviation of the nearest-neighbor exchange constants from the uniform interactions limit. In the absence of dimerization the AAHM is integrable with the Bethe ansatz \cite{Bethe31}, however an infinitesimal dimerization breaks integrability. At the opposite end of full dimerization the model is trivially integrable, with the chain breaking up in non-interacting dimers. Tuning the dimerization strength allows control of the proximity to integrability, in analogy with the next-nearest neighbor interaction in the one-dimensional chain \cite{NPK15}. To investigate the effect of integrability breaking on the rate of thermalization we calculate the time evolution of local correlations after an interacting quench, and more specifically their deviation from their thermal ensemble value as a function of time. We also calculate the difference between the diagonal and thermal ensemble values for local operators for increasing system size. This difference should go to zero in the thermodynamic limit (TDL) in the case of thermalization. In both calculations the dependence on the dimerization parameter is examined. As in the case of the AAHM with next-nearest neighbor interactions, we invoke the distribution of the operator off-diagonal elements with respect to the energy difference to determine the relative strength of temporal fluctuations for varying dimerization.


An infinitesimal dimerization breaks the integrability of the one-dimensional uniform AAHM and reduces its spatial symmetry by a factor of two by doubling the unit cell. This is qualitatively different with respect to the breaking of integrability by the addition of next-nearest neighbor interactions, which in contrast preserve the spatial symmetry of the Hamiltonian. In the latter case the operator off-diagonal elements that control the strength of temporal fluctuations of the operator's expectation value change smoothly as integrability is broken, and the time evolution takes place in the same symmetry subsector as in the integrable case. On the other hand, dimerization mixes symmetry subsectors of the uniform AAHM due to the spatial symmetry reduction. This means that even an infinitesimal dimerization changes qualitatively the form of the operator off-diagonal elements, as levels very close in energy immediately develop considerable overlaps. Such strong off-diagonal matrix elements between states close in energy are absent in the non-dimerized limit, and are responsible for the fastest thermalization occuring for very weak dimerization. In contrast, the AAHM with next-nearest neighbor interactions requires a stronger finite integrability breaking to generate significant overlaps between eigenstates not differing much in energy, and therefore to cause faster thermalization \cite{NPK15}.

In the opposite limit of full dimerization the AAHM is trivially integrable, with each non-zero bond connecting two spins to form an isolated dimer. For strong but not full dimerization the isolated dimers start to interact and form energy bands. Their interaction is however considerably weaker than intra-dimer exchange, consequently the probability flow between neighboring dimers is weak and thermalization is slow. This can also be seen in the off-diagonal elements of local operators between different eigenstates of the Hamiltonian. These operators act on isolated or neighboring dimers of the fully dimerized limit, which means that
they will have non-zero off-diagonal elements between eigenstates whose energies differ at the level of one or two dimers and are therefore quite close.

According to the above, the breaking of integrability of the dimerized AAHM is qualitatively different in the two extremes of weak and strong dimerization, as in the former case it assists thermalization in a much more efficient way due to the immediate reduction of spatial symmetry. In the latter case the evolution away from the fully dimerized limit is smooth and efficient thermalization settles in more slowly, as in the case of the AAHM with next-nearest neighbor interactions. It is found that the time-dependent local operator expectation values evolve smoothly between the non-dimerized and the fully dimerized limit, meaning that thermalization is getting slower as the dimerization is getting stronger. Further evidence for this is provided from the operator off-diagonal elements, which also evolve with the dimerization smoothly between the two extreme limits, and also from the difference between the diagonal and thermal ensemble values. The result is that for moderate dimerization thermalization is already quite slower. Here the interaction term of the AAHM is chosen comparable in magnitude to the tunneling term, with the interaction also stronger than the dimerization. Similar results for considerable dimerization have been found when the interaction term is weaker than the tunneling term \cite{Essler14}, where the slow approach to thermalization was associated with the emergence of the prethermalized regime.


The plan of this paper is as follows: Section \ref{sec:Model} introduces the model and the methods to be used. Section \ref{sec:timeevolution} discusses the time evolution of an operator expectation value after a quench and the requirements for thermalization. Section \ref{subsec:scalingcorrelations} includes the results, namely time-dependent local operator expectation values and the scaling of the difference between the diagonal and canonical thermal ensemble values for an interacting quench. It also presents the distribution of the strength of local operator off-diagonal terms with energy difference. Finally Sec. \ref{sec:conclusions} presents the conclusions.

\section{Model and Method}
\label{sec:Model}

The Hamiltonian of the dimerized AAHM model for a chain is
\begin{eqnarray}
H & = & J \sum_{i=1}^N [ ( 1 - (-1)^i \delta ] ( s_i^x s_{i+1}^x + s_i^y s_{i+1}^y + \Delta s_i^z s_{i+1}^z ) \textrm{ }
\label{eq:Hchain}
\end{eqnarray}

with $N$ the number of spins. The spin magnitude $s=1/2$ and the boundary conditions are periodic so that $s_{N+1} \equiv s_1$. In contrast to the non-dimerized case the unit cell contains two spins.
In order for the ground state before the quench and consequently the time-evolved wavefunction to be included in the one-dimensional even irreducible representation \cite{Altmann94}, the strong and weak bonds must both appear an even number of times, therefore $N$ is taken to be a multiple of four. The dimerization parameter $\delta$ is taken to be positive and breaks the integrability of Hamiltonian (\ref{eq:Hchain}) once it is non-zero, with the exchange constants alternating as $J(1 \pm \delta)$. Hamiltonian (\ref{eq:Hchain}) is trivially integrable in the fully dimerized limit $\delta=1$.
$J$ is taken to be 1 from now on, defining the unit of energy.

To calculate diagonal and thermal ensemble values Hamiltonian (\ref{eq:Hchain}) has to be fully diagonalized, and in order to do so its symmetries are exploited \cite{NPK05,NPK07,NPK09,NPK15-1,NPK04}. The full symmetry group of the Hamiltonian is the product of the spatial and spin symmetry groups. The spatial group is the dihedral $D_{\frac{N}{2}}$ group \cite{Altmann94}, while in spin space the Hamiltonian is symmetric under inversion when $S^z=0$.

To calculate directly the time evolution of operators the Chebyshev polynomial expansion is employed \cite{Zhang07,Zhang08}. It is more efficient in comparison with exact diagonalization for this purpose, making it possible to consider larger system sizes. In addition, the block-diagonalization of Hamiltonian (\ref{eq:Hchain}) according to the irreducible representations of its total symmetry group allows to time-evolve the initial wavefunction separately within each representation, increasing the maximum system size that can be considered even more. If the initial wavefunction is confined to a single symmetry subsector, it evolves in time only in this subsector and furthermore the diagonal ensemble has non-zero contributions only from expectation values of this subsector. This is the case for the interacting quench considered in this paper. The initial state is the ground state of Hamiltonian (\ref{eq:Hchain}) before the quench.
For the initial state different values of $\Delta_0$ are chosen for different $\delta$ so that the post-quench effective temperature is equal to $\beta=1.3500$ in every case. After the quench $\Delta=1.1$ is taken.

In the absence of dimerization ($\delta=0$) Hamiltonian (\ref{eq:Hchain}) is integrable via the Bethe ansatz \cite{Bethe31}, and there is quasi-long range order in the ground state. Once $\delta \neq 0$ a gap opens and the ground state has N\'eel order \cite{Chitra95,Goli13}. Here we consider relatively small $\delta \leq 0.3$, and investigate how integrability breaking affects thermalization by considering the nearest-neighbor operators $s_j^x s_k^x + s_j^y s_k^y$ and $s_j^z s_k^z$, on the strong ($j=2i+1$, $k=2i+2$) and weak ($j=2i$, $k=2i+1$) bonds. We pick $\Delta=1.1$ for the anisotropy parameter after the quench to rule out mixing of different $S$ sectors, which can occur in the spin isotropic SU(2) case when one works in the $S^z$ basis.

\section{Time Evolution}
\label{sec:timeevolution}
The initial wavefunction $\vert \Psi(t=0) \rangle = \vert \Psi(0) \rangle$ is the ground state of the Hamiltonian before the quench. If the post-quench eigenvalues and
eigenvectors of Hamiltonian (\ref{eq:Hchain}) are $E_n$ and
$\vert \Psi_n \rangle$, it is for the wavefunction $\vert \Psi (t) \rangle$
after the quench 
\begin{eqnarray}
\vert \Psi (t) \rangle = e^{-iHt} \vert \Psi(0) \rangle = \sum_n C_n e^{-iE_nt} \vert \Psi_n \rangle
\end{eqnarray}
The coefficients $C_n \equiv \langle \Psi_n \vert \Psi(0) \rangle$ do not depend on time. The time evolution takes place in the one-dimensional even irreducible representation (Sec. \ref{sec:Model}), therefore there are no degeneracies and an operator $\hat{O}$ at time $t$ has the expectation value
\begin{eqnarray}
<\hat{O}(t)> & = & \sum_n \vert C_n \vert ^2 \langle \Psi_n \vert \hat{O} \vert \Psi_n \rangle + \nonumber \\ & & \sum_{m \neq n} C_n^* C_m e^{-i(E_m-E_n)t} \langle \Psi_n \vert \hat{O} \vert \Psi_m \rangle
\label{eq:diagonal}
\end{eqnarray}
In the time average of Eq. (\ref{eq:diagonal}) only the first term has non-zero contribution. The average is given by $\bar{O} \equiv \sum_n \vert C_n \vert ^2 \langle \Psi_n \vert \hat{O} \vert \Psi_n \rangle$, defining the value of $\hat{O}$ in the diagonal ensemble \cite{Rigol08}. The wavefunction evolves in time in the non-degenerate subsector of the ground state before the quench, therefore only eigenstates of this subsector have non-zero coefficients.
Thermalization requires the diagonal and canonical thermal ensemble values to be equal in the TDL. According to Eq. (\ref{eq:diagonal}) the strength of fluctuations around the diagonal ensemble value is controlled by the off-diagonal terms $\langle \Psi_n \vert \hat{O} \vert \Psi_m \rangle \equiv \hat{O}_{nm}$, $m \neq n$. Thermalization also requires that fluctuations with respect to $\bar{O}$ are small in the TDL. To determine the canonical thermal ensemble $\rho_{th}$ describing the equilibrated long-time limit, the energy
and the thermal average
are set equal to each other
\begin{eqnarray}
\langle \Psi(0) \vert H \vert \Psi(0) \rangle = Tr(H \rho_{th})
\end{eqnarray}
The numerical solution of this equation gives the temperature.

\section{Nearest-Neighbor Correlations}
\label{subsec:scalingcorrelations}

\begin{figure}
\includegraphics[width=3.5in,height=2.6in]{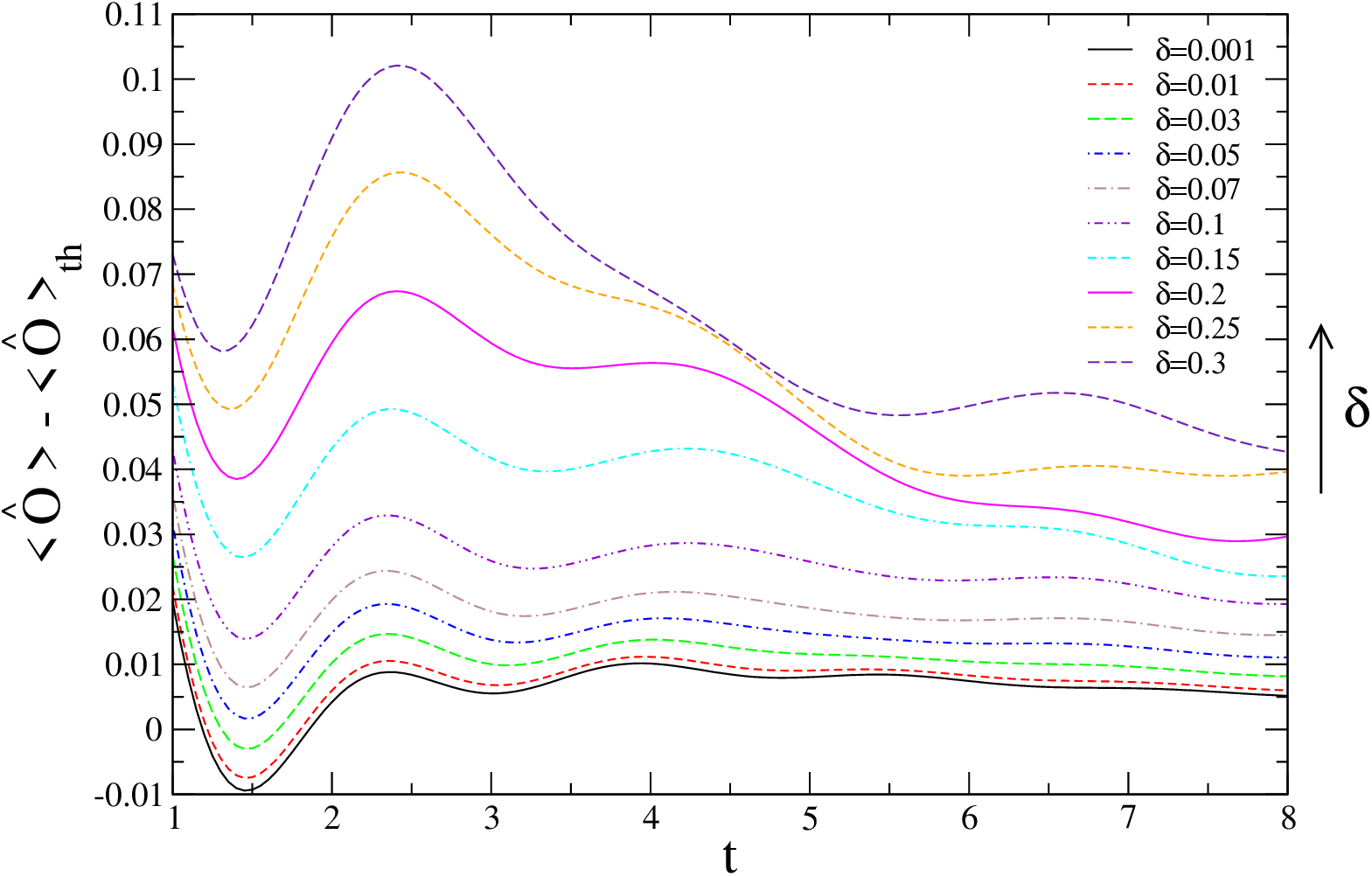}
\vspace{0pt}
\caption{(Color online) Time evolution of the expectation value of $\hat{O} = s_{2i+1}^x s_{2i+2}^x + s_{2i+1}^y s_{2i+2}^y$ with respect to its thermal value $<\hat{O}>_{th}$ for different values of $\delta$ with $\Delta=1.1$. Spins $s_{2i+1}$ and $s_{2i+2}$ interact via a strong bond $1+\delta$. For the initial state different values of $\Delta_0$ are chosen for different $\delta$ so that the post-quench effective temperature is equal to $\beta=1.3500$ for every $\delta$. Starting with the weakest and ending with the strongest $\delta$ the values of $\Delta_0$ are 88.94, 88.236, 82.95, 74.152, 64.1131, 50.124, 33.4604, 23.7088, 18.0157, and 14.5974.
}
\label{fig:sxy0sxy1time}
\end{figure}

To investigate thermalization the time evolution of local operator expectation values is of interest, therefore nearest-neighbor interactions are considered, which relate to a subsystem of two adjacent spins. The full correlation $\vec{s}_i \cdot \vec{s}_{i+1}$ is not calculated, as it commutes with Hamiltonian (\ref{eq:Hchain}) at the integrable points and for isotropic spin interactions $\Delta=1$. Instead correlations in the spin $xy$ plane and along the spin $z$ axis are considered. Fig. \ref{fig:sxy0sxy1time} shows the time evolution of the expectation value of the planar spin correlation $s_{2i+1}^x s_{2i+2}^x + s_{2i+1}^y s_{2i+2}^y$ between spins interacting via the strong exchange coupling $1 + \delta$. More accurately, it shows the deviation of the time-dependent expectation value from the canonical thermal ensemble value. Since the chains considered are of finite size, revivals of expectation values are eventually generated in later times. The results presented are for a $N=32$ chain, which generates the infinite chain results without any revivals for times at least up to $\frac{8}{J}$ \cite{Essler14,NPK15}. This has also been vindicated by comparing the time-dependent expectation values as a function of $N$, as they agree to progressively longer times with increasing $N$. The deviation from zero in Fig. \ref{fig:sxy0sxy1time} increases with the dimerization, indicating that thermalization becomes slower with $\delta$. Further evidence of that comes from the fluctuations for higher dimerizations, which are still of considerable strength for the highest times.

\begin{figure}
\includegraphics[width=3.5in,height=2.6in]{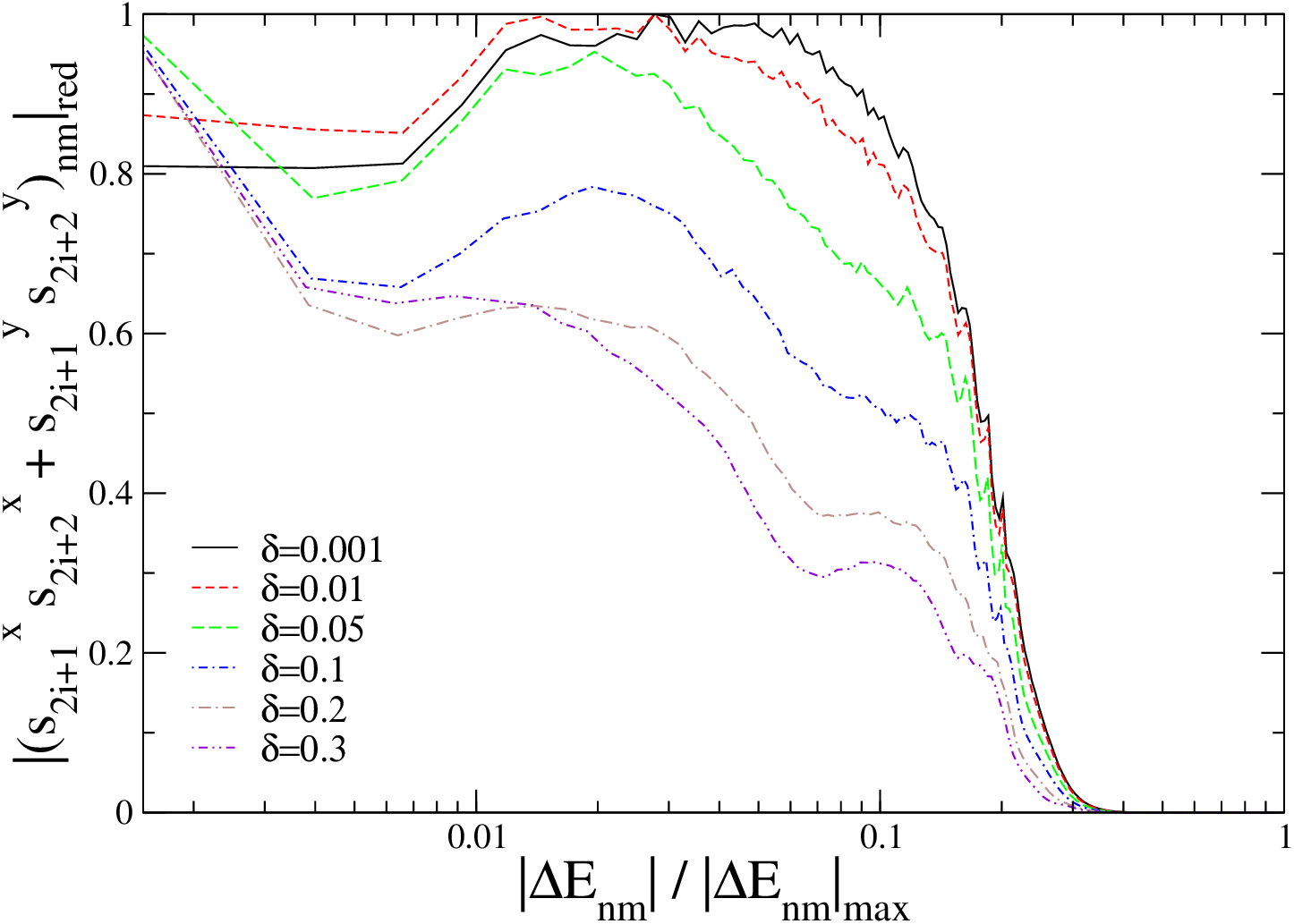}
\vspace{0pt}
\caption{(Color online) Distribution of matrix elements $(s_{2i+1}^x s_{2i+2}^x + s_{2i+1}^y s_{2i+2}^y)_{nm}$ as a function of the energy difference $\Delta E_{nm} = E_n-E_m$ for different values of $\delta$ with $\Delta=1.1$. What is plotted is the coarse grained average of $|(s_{2i+1}^x s_{2i+2}^x + s_{2i+1}^y s_{2i+2}^y)_{nm}|$ divided with its maximum value as a function of the reduced average energy difference for 100 bins with an equal number of points. Spins $s_{2i+1}$ and $s_{2i+2}$ interact via a strong bond $1+\delta$.
}
\label{fig:sxy0sxy1offdiagonal}
\end{figure}

It is instructive to compare the case of the dimerized Hamiltonian (\ref{eq:Hchain}) with the integrability breaking effected by adding uniform couplings between next-nearest neighbors. In this case thermalization becomes faster with the strength of the integrability breaking term \cite{NPK15}. This was explained by considering the energy distribution of the off-diagonal elements of the local operators, which according to Eq. (\ref{eq:diagonal}) control the strength of the fluctuations around the long-time limit. It was found that for very small next-nearest neighbor exchange the off-diagonal elements peak away from the origin. As the next-nearest neighbor exchange increases, a peak close to zero energy develops that becomes important in the long-time limit and facilitates thermalization. This is because the high-frequency fluctuations related to the higher-energy peak will eventually cancel each other out, and the low-energy peak will determine the fluctuations in the long-time limit, which will be slowly varying. Fig. \ref{fig:sxy0sxy1offdiagonal} shows the same calculation for the off-diagonal elements of $s_{2i+1}^x s_{2i+2}^x + s_{2i+1}^y s_{2i+2}^y$ for the post-quench Hamiltonian. In the integrable limit $\delta=0$ the Hamiltonian corresponds to the AAHM with zero next-nearest neighbor interactions, where the local operator off-diagonal elements are very small close to zero and have a maximum away from it \cite{NPK15}. For very small $\delta=0.001$ their distribution is quite different. Matrix elements between states very close in energy have now values comparable but smaller in magnitude than the higher-energy peak. Even though there is no well separated low-energy peak as in the case of the next-nearest neighbor AAHM, the present distribution also points to thermalization in the long-time limit in accordance with Fig. \ref{fig:sxy0sxy1time}, where the low-frequency fluctuations will eventually overtake the high-frequency ones.

The introduction of dimerization breaks the integrability of Hamiltonian (\ref{eq:Hchain}) and alters significantly the distribution of the off-diagonal elements. The time-evolved wavefunction belongs to a specific irreducible representation, which is determined by the ground state before the quench. In the integrable limit $\delta=0$, eigenstates close in energy that belong to the irreducible representation that solely includes the time-evolved wavefunction have local operator off-diagonal elements close to zero (see discussion in Ref. \cite{NPK15}). An infinitesimal $\delta$ reduces the spatial symmetry of the Hamiltonian by a factor of two by doubling the unit cell, and the spatial symmetry group changes from $D_N$ to $D_{\frac{N}{2}}$ \cite{Altmann94}.
The total number of irreducible representations is less for the $D_{\frac{N}{2}}$ group,
and their eigenstates are combinations of eigenstates of different irreducible representations of the $D_N$ group. There is no restriction on the strength of the off-diagonal elements of local operators between eigenstates with similar energies that belong to different irreducible representations of the $D_N$ group. The appearance of significant local operator off-diagonal elements between eigenstates close in energy for infinitesimal dimerization is thus not a perturbative effect, but originates in the reduction of the Hamiltonian spatial symmetry, which also results in the breaking of integrability. This is seen in Fig. \ref{fig:sxy0sxy1offdiagonal} for $\delta=0.001$ and 0.01.

In the opposite limit of strong dimerization Hamiltonian (\ref{eq:Hchain}) is trivially integrable for $\delta=1$, with each non-zero bond connecting two spins to form an isolated dimer. The eigenstates of the Hamiltonian are products of dimer eigenstates, with their eigenenergies being sums of dimer energies, which are $-\frac{3}{4}$ and $\frac{1}{4}$, scaled with the intra-dimer exchange interaction. For strong dimerization $\delta \lesssim 1$ the isolated dimers start to interact and form narrow energy bands around the discrete levels of the fully dimerized limit. The inter-dimer interaction is however considerably weaker than the intra-dimer exchange, consequently the probability flow between neighboring dimers is weak, which points to slow thermalization. This is in agreement with Fig. \ref{fig:sxy0sxy1time}, where the deviation from zero is considerably stronger for larger $\delta$ and also fluctuations are strong.
In addition, local operators act on approximately isolated dimers or weakly connected neighboring dimers, therefore they will only have non-zero off-diagonal elements between eigenstates whose energies differ at the level of one or two dimers, and are therefore close and belong to the same energy band.
This is seen in Fig. \ref{fig:sxy0sxy1offdiagonal}, where for the higher $\delta$'s the off-diagonal elements have a maximum for an energy difference close to zero, and decrease with increasing energy difference. This off-diagonal elements distribution is characteristic of a strongly dimerized Hamiltonian, with the strong low-frequency fluctuations leading to slow thermalization. Fig. \ref{fig:sxy0sxy1offdiagonal} also shows that the evolution of the off-diagonal elements distribution of $s_{2i+1}^x s_{2i+2}^x + s_{2i+1}^y s_{2i+2}^y$ in the eigenstate basis is smooth as a function of $\delta$ between the weak and strong dimerization limits. This points to slower thermalization for increasing dimerization, in agreement with Fig. \ref{fig:sxy0sxy1time}.

\begin{figure}
\includegraphics[width=3.5in,height=2.6in]{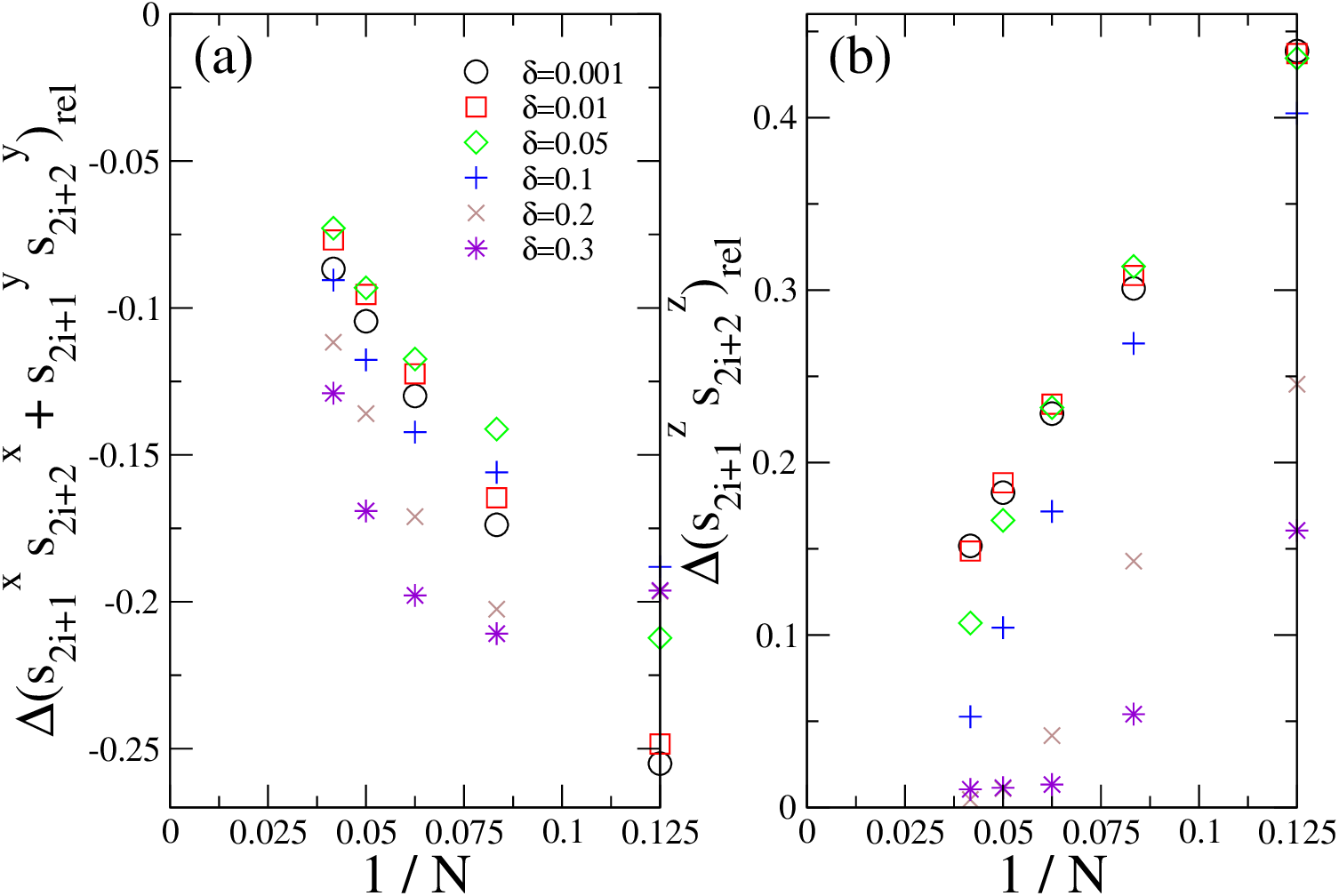}
\vspace{0pt}
\caption{(Color online) Scaling of the difference $\Delta\hat{O}_{rel}=\frac{\bar{O} - \langle \hat{O} \rangle _{th}}{\langle \hat{O} \rangle _{th}}$ between the diagonal and canonical thermal ensemble values divided by the canonical thermal ensemble value as a function of inverse length $\frac{1}{N}$ for correlation function (a) $\hat{O} = s_{2i+1}^x s_{2i+2}^x + s_{2i+1}^y s_{2i+2}^y$ and (b) $\hat{O} = s_{2i+1}^z s_{2i+2}^z$ for different values of $\delta$ with $\Delta=1.1$. Spins $s_{2i+1}$ and $s_{2i+2}$ interact via a strong bond $1+\delta$. The maximum chain has $N=24$. The post-quench effective temperature is equal to $\beta=1.3500$ for every $\delta$, and the values of $\Delta_0$ are given in the caption of Fig. \ref{fig:sxy0sxy1time}.
}
\label{fig:s0s1scaling}
\end{figure}

Further evidence for the dependence of the thermalization rate on $\delta$ comes from the scaling of the difference between the diagonal and the canonical thermal ensemble values with $N$. This difference scaled with the canonical thermal ensemble value is shown in Fig. \ref{fig:s0s1scaling}(a). For the two smallest $\delta$ values the dependence on $\frac{1}{N}$ appears very close to linear. If a straight line is fitted through the data, the intercept is equal to $3.6 \times 10^{-3}$ for $\delta=0.001$ and $-7.2 \times 10^{-3}$ for $\delta=0.01$. This shows that chains of maximum size $N=24$ point to thermalization in the TDL for weak dimerization, with thermalization being faster for smaller $\delta$. For the next highest $\delta=0.05$ the line is significantly curved, and even a fit through the last two points does not pass close from the origin. This points to slower thermalization in agreement with Fig. \ref{fig:sxy0sxy1offdiagonal}, where for $\delta=0.05$ the low and the high-energy peak become approximately of the same strength. For even higher $\delta$ the lines are also curved, and also the distribution of the off-diagonal elements has its highest weight close to zero, both indicating even slower equilibration to a long-time limit as the Hamiltonian gets even closer to the strongly dimerized limit. Due to the lack of quicker thermalization one would have to go to higher $N$ to deduce the dependence of the ensemble difference on $\frac{1}{N}$.

\begin{figure}
\includegraphics[width=3.5in,height=2.6in]{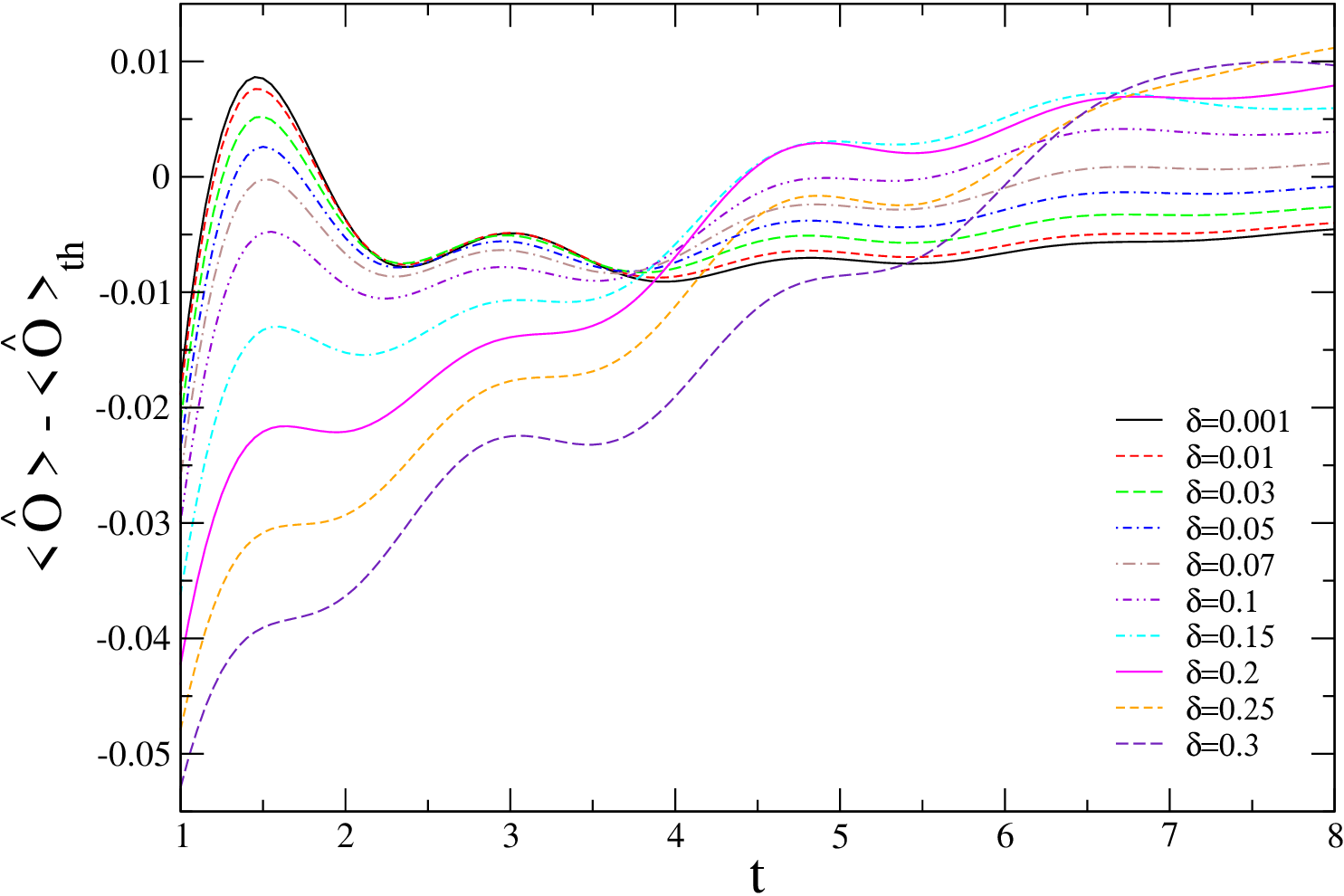}
\vspace{0pt}
\caption{(Color online) Time evolution of the expectation value of $\hat{O} = s_{2i+1}^z s_{2i+2}^z$ with respect to its thermal value $<\hat{O}>_{th}$ for different values of $\delta$ with $\Delta=1.1$. Spins $s_{2i+1}$ and $s_{2i+2}$ interact via a strong bond $1+\delta$. The post-quench effective temperature is equal to $\beta=1.3500$ for every $\delta$, and the values of $\Delta_0$ are given in the caption of Fig. \ref{fig:sxy0sxy1time}.
}
\label{fig:sz0sz1time}
\end{figure}

\begin{figure}
\includegraphics[width=3.5in,height=2.6in]{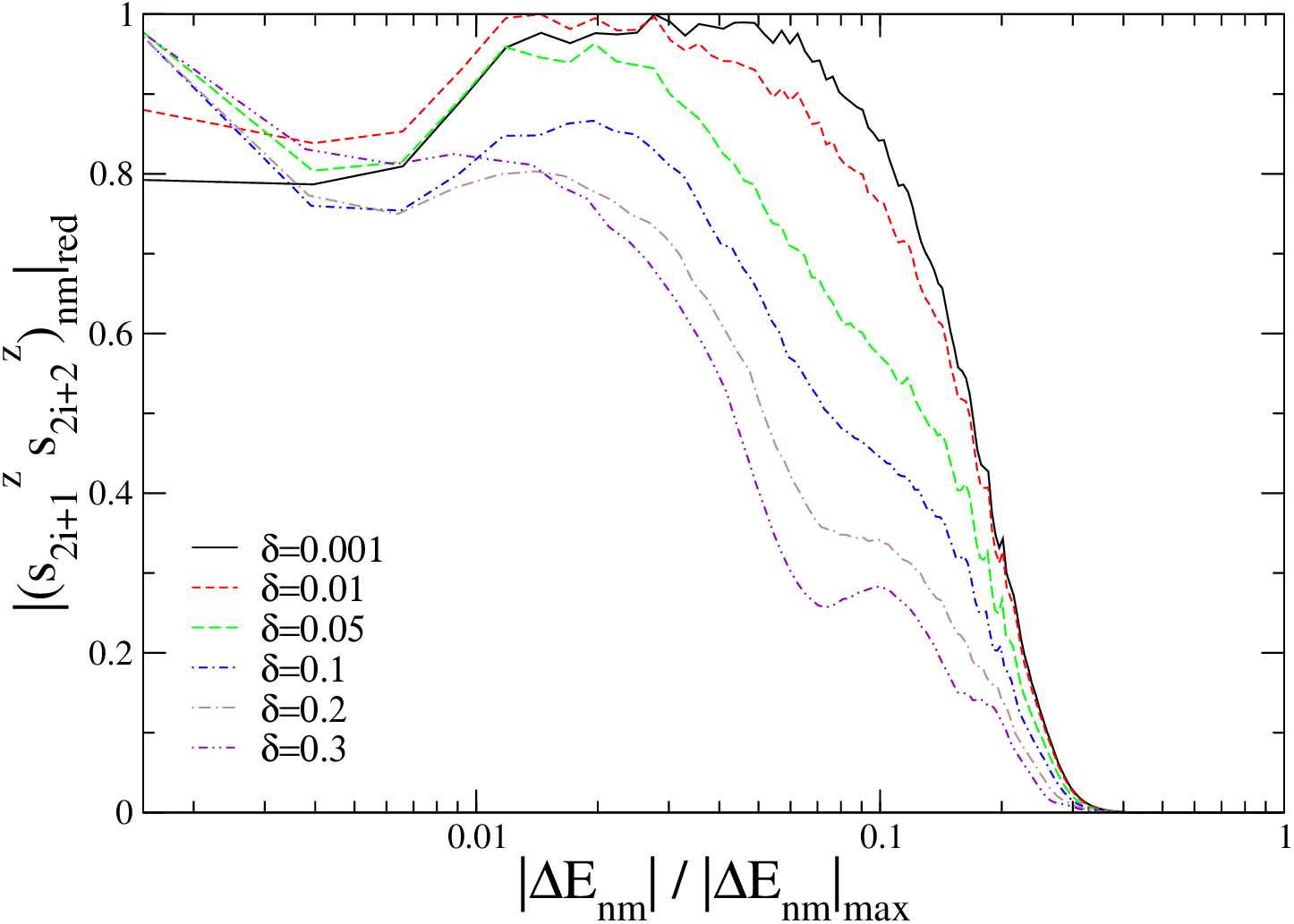}
\vspace{0pt}
\caption{(Color online) Distribution of matrix elements $(s_{2i+1}^z s_{2i+2}^z)_{nm}$ as a function of the energy difference $\Delta E_{nm} = E_n-E_m$ for different values of $\delta$ with $\Delta=1.1$. What is plotted is the coarse grained average of $|(s_{2i+1}^z s_{2i+2}^z)_{nm}|$ divided with its maximum value as a function of the reduced average energy difference for 100 bins with an equal number of points. Spins $s_{2i+1}$ and $s_{2i+2}$ interact via a strong bond $1+\delta$.
}
\label{fig:sz0sz1offdiagonal}
\end{figure}

Fig. \ref{fig:s0s1scaling}(b) shows the corresponding result for correlation function $s_{2i+1}^z s_{2i+2}^z$. A similar conclusion can be drawn. The intercepts for a straight line fit through the data for $\delta=0.001$ and 0.1 are 0.011 and 0.016 respectively, indicating faster thermalization for smaller $\delta$, while for higher $\delta$ the dependence on $\frac{1}{N}$ becomes again non-linear, at least for the available sizes. In Fig. \ref{fig:sz0sz1time} the difference of the time-dependent expectation value of $s_{2i+1}^z s_{2i+2}^z$ from its thermal ensemble value is plotted as a function of time. $\delta=0.001$ is not the closest curve to zero for the longest available time, however for higher $\delta$ the deviation has already changed sign and this is also expected for the deviations for $\delta$ closer to 0.001. This conclusion is also supported by the scaling data of Fig. \ref{fig:s0s1scaling}(b), which resemble a straight line best for weak $\delta$, and also from the distribution of the off-diagonal elements, which is shown in Fig. \ref{fig:sz0sz1offdiagonal} and is similar to Fig. \ref{fig:sxy0sxy1offdiagonal}. Again the results for correlation function $s_{2i+1}^z s_{2i+2}^z$ support the thermalization scenario where stronger dimerization slows down thermalization.

The fluctuations in Fig. \ref{fig:sz0sz1time} are the strongest for higher $\delta$. Fluctuations for larger $\delta$ are expected to be significant, judging also from the distributions of Fig. \ref{fig:sz0sz1offdiagonal}. The strong fluctuations of the difference around zero for the higher $\delta$ lead to a relative difference in Fig. \ref{fig:s0s1scaling}(b) that approaches zero fast with system size, even though the subsystem has not thermalized yet. This is something that does not occur for $s_{2i+1}^x s_{2i+2}^x + s_{2i+1}^y s_{2i+2}^y$ or in the AAHM with nearest and next-nearest neighbor interactions \cite{NPK15}. The reason is that in these cases the difference does not change sign for later times.

\begin{figure}
\includegraphics[width=3.5in,height=2.6in]{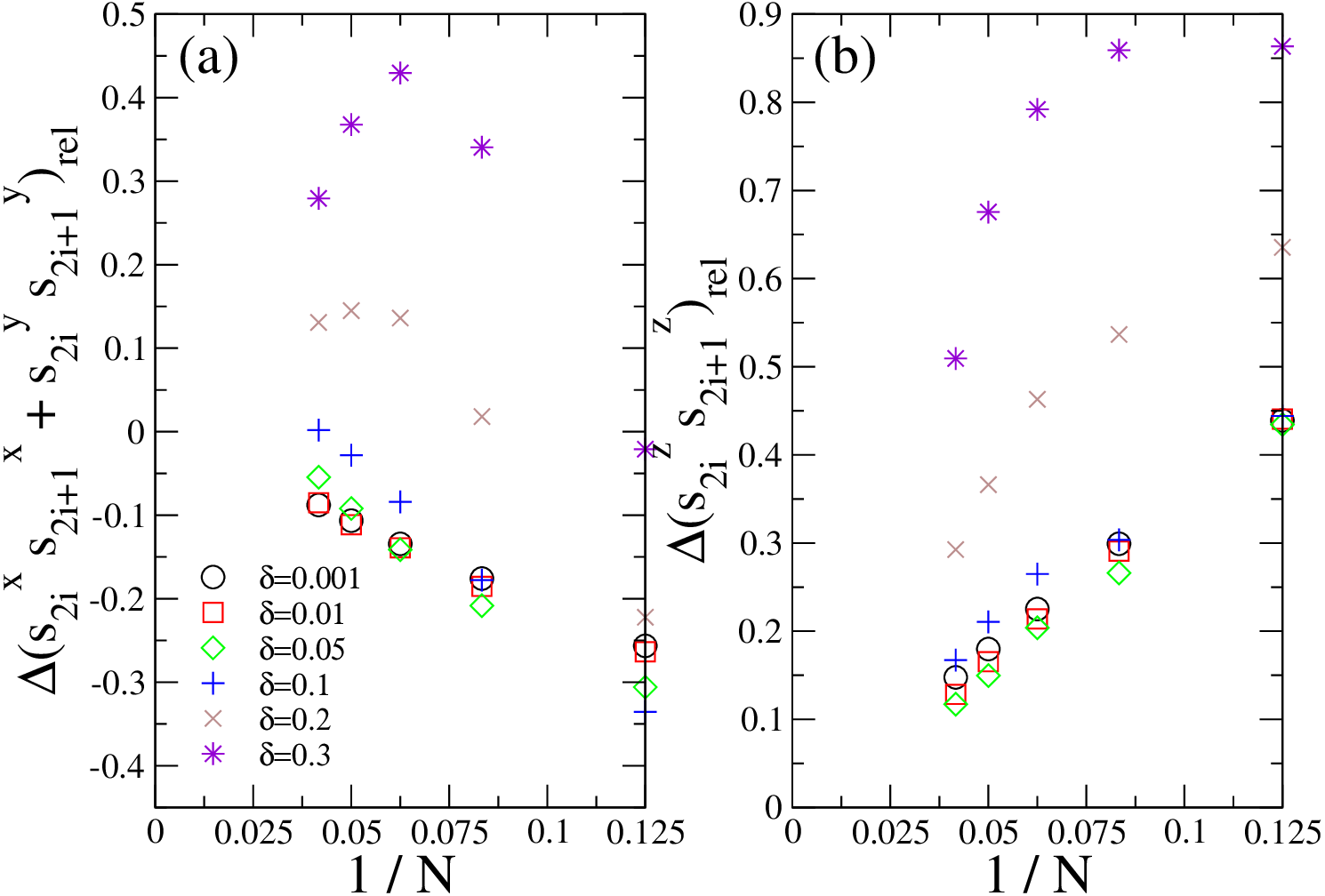}
\vspace{0pt}
\caption{(Color online) Scaling of the difference $\Delta\hat{O}_{rel}=\frac{\bar{O} - \langle \hat{O} \rangle _{th}}{\langle \hat{O} \rangle _{th}}$ between the diagonal and canonical thermal ensemble values divided by the canonical thermal ensemble value as a function of inverse length $\frac{1}{N}$ for correlation functions (a) $\hat{O} = s_{2i}^x s_{2i+1}^x + s_{2i}^y s_{2i+1}^y$ and (b) $\hat{O} = s_{2i}^z s_{2i+1}^z$ for different values of $\delta$ with $\Delta=1.1$. Spins $s_{2i}$ and $s_{2i+1}$ interact via a weak bond $1-\delta$. The maximum chain has $N=24$. The post-quench effective temperature is equal to $\beta=1.3500$ for every $\delta$, and the values of $\Delta_0$ are given in the caption of Fig. \ref{fig:sxy0sxy1time}.
}
\label{fig:s1s2scaling}
\end{figure}

\begin{figure}
\includegraphics[width=3.5in,height=2.6in]{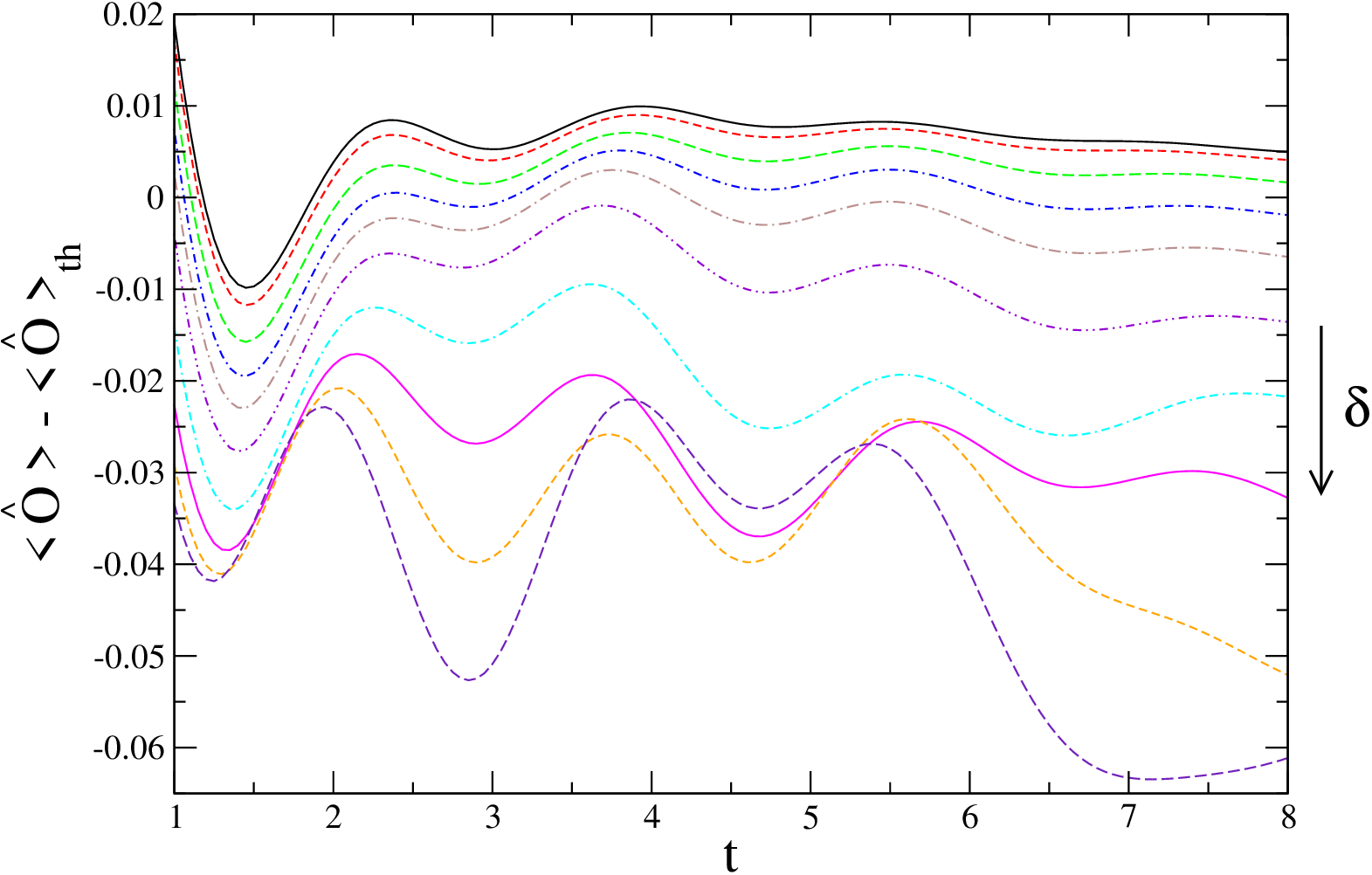}
\vspace{0pt}
\caption{(Color online) Time evolution of the expectation value of $\hat{O} = s_{2i}^x s_{2i+1}^x + s_{2i}^y s_{2i+1}^y$ with respect to its thermal value $<\hat{O}>_{th}$ for different values of $\delta$ with $\Delta=1.1$. Spins $s_{2i}$ and $s_{2i+1}$ interact via a weak bond $1-\delta$. The labels for the different lines are the same as in Figs. \ref{fig:sxy0sxy1time} and \ref{fig:sz0sz1time}. The post-quench effective temperature is equal to $\beta=1.3500$ for every $\delta$, and the values of $\Delta_0$ are given in the caption of Fig. \ref{fig:sxy0sxy1time}.
}
\label{fig:sxy1sxy2time}
\end{figure}

\begin{figure}
\includegraphics[width=3.5in,height=2.6in]{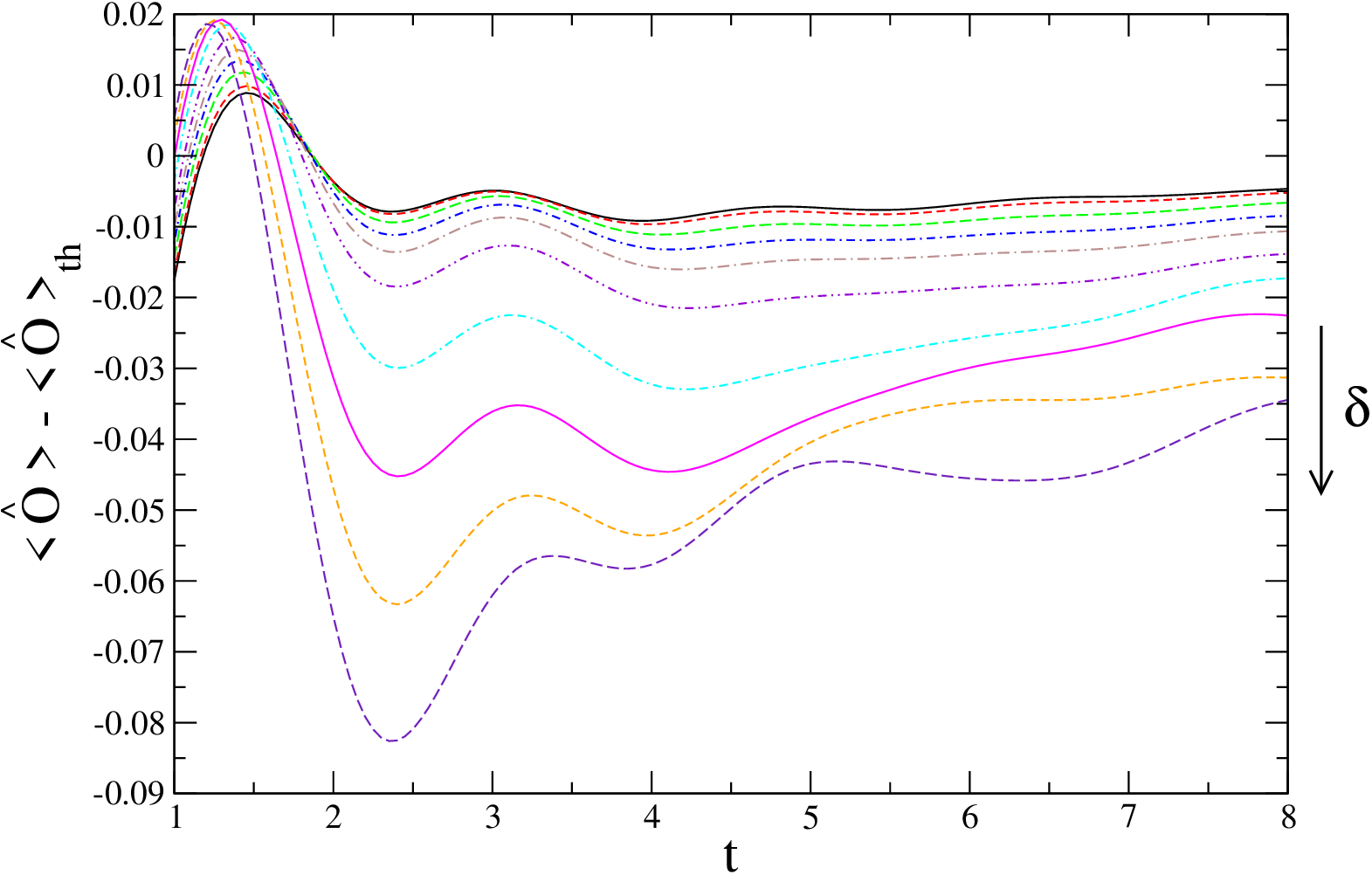}
\vspace{0pt}
\caption{(Color online) Time evolution of the expectation value of $\hat{O} = s_{2i}^z s_{2i+1}^z$ with respect to its thermal value $<\hat{O}>_{th}$ for different values of $\delta$ with $\Delta=1.1$. Spins $s_{2i}$ and $s_{2i+1}$ interact via a weak bond $1-\delta$. The labels for the different lines are the same as in Figs. \ref{fig:sxy0sxy1time} and \ref{fig:sz0sz1time}. The post-quench effective temperature is equal to $\beta=1.3500$ for every $\delta$, and the values of $\Delta_0$ are given in the caption of Fig. \ref{fig:sxy0sxy1time}.
}
\label{fig:sz1sz2time}
\end{figure}

Similar conclusions for slower thermalization with increasing $\delta$ are drawn from the correlation function $s_{2i}^x s_{2i+1}^x + s_{2i}^y s_{2i+1}^y$ between nearest-neighbor spins interacting via the weak exchange coupling $1 - \delta$. Fig. \ref{fig:s1s2scaling}(a) shows the scaling data, where for the smallest $\delta$ values 0.001 and 0.01 a straight line fit passes very close from the origin. For higher $\delta$ this is not the case, and the dependence on $\frac{1}{N}$ deviates from a straight line, at least for the available $N$. Fig. \ref{fig:sxy1sxy2time} shows the time evolution of the expectation value, which is similar to the one of $s_{2i+1}^z s_{2i+2}^z$ and also points to slower thermalization with increasing dimerization. In Fig. \ref{fig:sxy1sxy2time} it is expected that for small $\delta$ the deviations will also change sign for longer times, as is also the expectation for Fig. \ref{fig:sz0sz1time}. The same thermalization pattern is suggested by correlation $s_{2i}^z s_{2i+1}^z$ from the results shown in Figs. \ref{fig:s1s2scaling}(b) and \ref{fig:sz1sz2time}. For both correlation functions on the weak bonds the thermalization predictions are supported by the distributions of the off-diagonal elements of the operators, which are not shown but are very similar to the ones of Figs. \ref{fig:sxy0sxy1offdiagonal} and \ref{fig:sz0sz1offdiagonal}.

\section{Conclusions}
\label{sec:conclusions}

It is generally expected that a non-integrable Hamiltonian thermalizes according to the canonical thermal ensemble. In this paper we considered a Hamiltonian that interpolates between two integrable points by varying the degree of dimerization of its nearest-neighbor interactions.
It was found that thermalization slows down with the dimerization strength. This was shown by three different types of calculations. First the direct time evolution of expectation values of local operators with time was computed, as well as the difference between the diagonal and thermal ensemble values with increasing system size for these operators. Finally the distribution of local operator off-diagonal elements as a function of energy, which control the temporal fluctuations, was calculated. These calculations show that the reduction of spatial symmetry leads to the quickest thermalization for infinitesimal dimerization. This contrasts the AAHM with nearest and integrability breaking next-nearest neighbor interactions, where a next-nearest neighbor interaction of significant strength is required to speed up thermalization \cite{NPK15}. In the fully dimerized limit the chain breaks up into isolated dimers which do not exchange probability, therefore there is no thermalization and the off-diagonal elements peak close to zero. The calculated quantities evolve smoothly between the non-dimerized and the fully dimerized limit and show how increasing dimerization slows down thermalization. A dimerized Hamiltonian is also important as it can be considered the first step on the way to a random Hamiltonian, where randomness can lead to a complete lack of thermalization \cite{Khatami12}.

The author is very thankful to A. Lazarides.

\bibliography{paperdimer}

\end{document}